\documentclass{vgtc}                          

\graphicspath{{figures/}{pictures/}{images/}{./}} 

\usepackage{times}                     

\usepackage{tabu}                      
\usepackage{booktabs}                  
\usepackage{lipsum}                    
\usepackage{mwe}                       
\usepackage{amsmath}
\usepackage{mathptmx}                  

\usepackage{xcolor}
\usepackage{soul}
\usepackage{wrapfig}

\newcommand{\ie}{{i.e.,}}
\newcommand{\eg}{{e.g.,}}

\newcommand{\ala}{{\`a la}}

\newcommand\chang[1]{{\color{black}#1}}

\newcommand{\secref}[1]{\hyperref[#1]{Sec.~\ref*{#1}}}
\newcommand{\appendixref}[1]{\hyperref[#1]{Appendix~\ref*{#1}}}
\newcommand{\figref}[1]{\hyperref[#1]{Fig.~\ref*{#1}}}
\newcommand{\eqnref}[1]{\hyperref[#1]{Eqn.~\ref*{#1}}}
\newcommand{\tabref}[1]{\hyperref[#1]{Table ~\ref*{#1}}}

\definecolor{quoteColor}{HTML}{ff5733}

\newcommand{\pxx}[1]
{\textbf{P}$_{\textrm{#1}}$}

\newcommand{\parahead}[1]
{%
  \paraheadd{#1}.}

\newcommand{\paraheadd}[1]
{%
  \vspace{0.07in}%
  \noindent%
  \textbf{\textit{#1}}%
}

\def\subsubsec#1
{\subsubsection{#1}}

\usepackage{tikz}
\usetikzlibrary{positioning}

\newcommand{\designcallout}[2]{%
\vspace{-0em}
\noindent
\begin{tikzpicture}
\node[
    draw=black,
    line width=1pt,
    rounded corners=8pt,
    inner xsep=4pt,
    inner ysep=0.8em,
    outer sep=0pt,
    text width=\dimexpr\linewidth-10pt-2pt\relax,
    align=left
] (box) {%
    \noindent{}\textit{#2}
};

\node[
    anchor=west,
    fill=white,
    inner xsep=5pt,
] (title) at ([xshift=14pt]box.north west) {\textsc{#1}};

\end{tikzpicture}%
\vspace{-0.5em}
}

\onlineid{5668}

\vgtccategory{Research}

\vgtcinsertpkg

\title{Where LLMs Fail with Visualization DSLs}

\author{
Chang Han\thanks{All authors are with University of Utah. \\
Emails: \{chang.han, andrew.mcnutt, kisaacs\}@sci.utah.edu} %
\and Andrew M. McNutt
\and Katherine E. Isaacs
     }

\teaser{
  \centering
  \includegraphics[width=\linewidth]{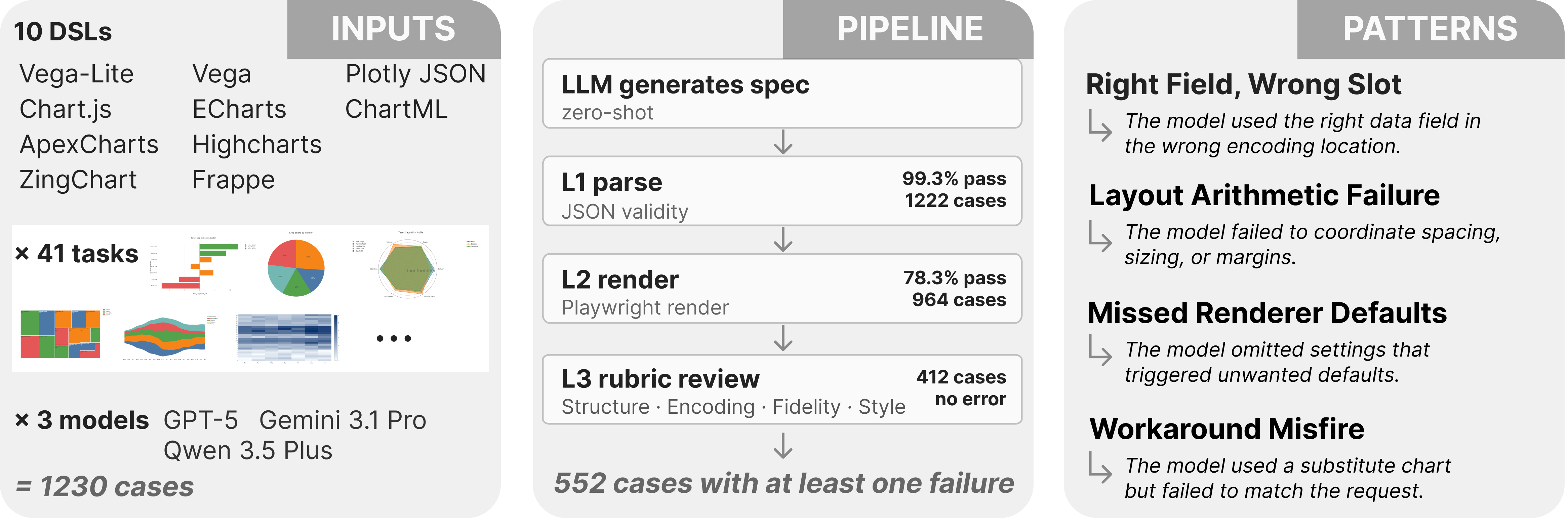}
  \caption{
  This study (left, middle) explores the patterns (right) that occur across DSL usage of visualization DSLs.
The patterns we observe show how mismatches between model assumptions and DSL conventions can result in LLMs producing invalid or incorrect visualizations---highlighting areas for future DSL design consideration.
  }
  \label{fig:teaser}
}

\abstract{  
As LLMs take up the role of authoring charts using visualization DSLs, the human constraints that shaped those languages may no longer apply----what is easy for a person is not necessarily easy for a model. 
To understand how LLMs might work better with DSLs, we explore where and how they fail with current DSL designs.
We evaluate 10 JSON-style visualization DSLs with 41 tasks across 3 LLMs, then assess the generated specifications with JSON and rendering checks, and qualitative coding of failed cases.
Analyzing how this specification generation process fails,
we identify four recurring failure patterns, link each to specific DSL features, and \chang{discuss design considerations for} future DSL designs.
}

\begin{document}

\maketitle

\section{Introduction}
LLM-based agents are increasingly being tasked to translate natural-language requests into chart specifications~\cite{dibia2023lida, wang2023data}, shifting the human role from coding to specifying intent and inspecting output.
\chang{Chart generation through DSLs is already embedded in diverse workflows---from interactive tools that delegate specification writing to AI~\cite{wang2025data} to programmatic toolkits that emit specifications for downstream applications~\cite{narechania2020nl4dv, zong2020lyra, mcnutt2021integrated}---where users increasingly inspect the rendered chart rather than the specification behind it.}
This shift challenges a fundamental assumption behind many visualization domain-specific languages (DSLs) and authoring tools: that the author is human.
Concise syntax, towers of defaults~\cite{wickham2010layered}, and hierarchical structure have long been designed~\cite{satyanarayan2016vega} around cognitive dimensions~\cite{green1989cognitive} such as  memory, attention, and effort. 
When the author is a language model, those trade-offs may change as 
what is easy for a human may be difficult for a model.

As a first step in understanding how language designs might shift, 
this paper studies how LLMs fail when authoring JSON-style visualization DSLs. 
We compare 10 DSLs across 41 visualization tasks, covering common authoring patterns such as single charts, layered views, facets, and composite charts. 
We focus on JSON-style DSLs because they are widely used in declarative visualization grammars~\cite{satyanarayan2016vega, li2018echarts} and share enough surface structure to make cross-DSL failure comparison meaningful.

Across \chang{1230 trials (1223 generated specifications)} from three LLMs, we identify four recurring failure patterns. In \textbf{\chang{encoding mismatch}}, models select the intended data field but place it in an incorrect position. In \textbf{layout arithmetic failure}, they fail to coordinate the many small spatial decisions needed for titles, legends, axes, and panels. In \textbf{missed renderer defaults}, they omit settings whose default behavior changes the chart. In \textbf{\chang{failed workaround}}, they approximate unsupported chart types with nearby primitives or custom-built compositions that do not preserve the requested semantics. 
%

These failures are not random syntax mistakes---they repeatedly trace back to DSL design choices. Based on these findings, \chang{we identify several design considerations for visualization DSLs that may work better with models. We suggest that} 
(1) DSLs should expose common visual encodings as explicit fields or channels rather than requiring placement inside nested containers; (2) DSLs should let composition primitives own ordinary layout budgeting instead of dispersing numeric controls across titles, legends, axes, and panels; and (3) DSLs should make renderer defaults and hidden contracts visible at authoring time.
AI-native visualization DSL design is not simply about shorter syntax or broader permissiveness, but about representing visual intent in forms whose structure and semantics LLMs can reliably interpret and compose. 

\section{Related Work}

Our work builds upon prior studies of visualization DSLs, as well as works exploring the interaction between LLMs and visualizations. 

\parahead{Visualization DSLs}
The Grammar of Graphics (GoG) tradition~\cite{wilkinson2011grammar, wickham2010layered} inspired a range of visualization grammars or languages~\cite{satyanarayan2014declarative, satyanarayan2016vega, li2018echarts}.
These include libraries and languages for general statistic charting~\cite{plotly_json_chart_schema, highcharts, chartjs, wongsuphasawat2020encodable}, as well as more specialized systems for  animation~\cite{kim2020gemini, ge2020canis} or annotation~\cite{rahman2025annogram, chen2025chartmark}.
\chang{\chang{Most recently, Flint~\cite{wang2026flint} introduced a semantics-driven intermediate language for both human authors and AI agents that compiles concise, library-agnostic specifications to multiple visualization backends.}}
McNutt~\cite{mcnutt2022no} surveyed those DSLs specifically centered in data serialization formats (like JSON), highlighting variations in specification structure (formal vs. colloquial), layout primitives, and chart-type vocabularies.
Existing evaluations have targeted human authoring trade-offs~\cite{satyanarayan2019critical, pu2023data} or quantitative design-pattern differences~\cite{kruchten2023metrics}; we instead probe how LLMs generate these DSLs, with consideration of the languages themselves.


\paraheadd{LLMs for Creating Visualization} A growing body of work studies LLM-driven chart generation, spanning end-to-end systems~\cite{dibia2023lida, narechania2020nl4dv, wang2025data, wang2023data, jobstintegrating} and natural language
to chart benchmarks~\cite{luo2021synthesizing, luo2021natural, chen2024viseval}. 
Some works bypass DSLs entirely by using LLMs to directly generate SVGs~\cite{JuanVector2025}, while chart-QA benchmarks address the dual problem of reading charts~\cite{zeng2025}. 
Existing authoring benchmarks~\cite{wang2024visualization, chen2024viseval} fix a single DSL---such as Vega-Lite or matplotlib---and vary the models and tasks. 
Outside visualization, code-generation evaluations~\cite{cassano2023multipl} and constrained-decoding work~\cite{scholak2021picard} show that language and schema choices materially shape LLM authoring; we ask the analogous question for visualization, varying the visualization DSL to examine failures as evidence about DSL design itself.
\chang{Notably, even agentic and code-generation pipelines often emit a declarative JSON specification as their intermediate representation~\cite{wang2023data, wang2025data, han2026slidesavr}, because specs are validatable, diffable, and safer to execute than arbitrary code---making the design of these DSLs relevant beyond direct authoring.}







\section{Study Design}

In this study (illustrated in \figref{fig:teaser}), we combine sets of DSLs, tasks, and models into a test suite of trials, evaluated for syntactic and  semantic correctness (via LLM-as-judge plus manual review).

\parahead{DSLs}
We started from the \chang{19} DSLs in the ``Charting'' category of McNutt's~\cite{mcnutt2022no} JSON-style visualization DSL survey, which covers tools for general statistical graphics\chang{, plus ZingChart, which the survey files under ``Responsive Vis'' but which is a general-purpose charting DSL (20 candidates)}.
From this initial set, we filtered to those that have a publicly installable runtime (excluding XML Charts~\cite{xml_swf_charts} for example), appear to be actively maintained (skipping \chang{the XML-based} ChartML~\cite{saito2009client}), and render headlessly in our \chang{Playwright~\cite{playwright} browser-automation} pipeline without proprietary tooling (forgoing FusionCharts~\cite{fusioncharts_fusionexport}). 

This yielded the following DSLs: Vega-Lite~\cite{satyanarayan2016vega}, Vega~\cite{satyanarayan2014declarative}, ECharts~\cite{li2018echarts}, 
Plotly JSON~\cite{plotly_json_chart_schema}, Highcharts~\cite{highcharts}, Chart.js~\cite{chartjs}, ApexCharts~\cite{apexcharts}, ZingChart~\cite{zingchart}, and Frappe~\cite{frappe_charts}. 
We also include a recent ChartML project~\cite{chartml}, distinct from 
the ChartML listed in McNutt's survey, as an actively maintained model-familiarity baseline, for a final set of 10 grammars.

\parahead{Tasks and Synthetic Data}
We use the generation of specific charts as our tasks, using an existing chart taxonomy as a basis, namely Data Viz Catalogue~\cite{dataviz_catalogue}, for its 
{extensive} scope and practical content (compared to collections such as Xenographics~\cite{xenographics}).
We then filtered out cases that would mainly test custom icons, freeform drawing, or bespoke layout algorithms rather than chart-spec writing. For instance, brainstorm or mind-map diagrams, word clouds, and Venn diagrams were excluded.
We then manually augmented the task set with cases useful for probing DSL behavior. These included composite tasks such as bar-line combinations, dual axes, reference lines, faceted views, and repeated small multiples, among others. 
Each task pairs a target chart type, plus additional constraints such as a required title, color mapping, field choice, legend, axis range, ordering rule, or explicit instruction not to use a decoy field. This design lets us test whether models preserve the requested data and presentation semantics, rather than only producing a plausible chart of the right general type.

\newcommand{\task}[1]{``\texttt{#1}''}

Pilot runs showed that models handled simple prompts (\eg{} \task{a bar chart of sales revenue}) very well.
To probe the boundaries of model capabilities, we used tasks with explicit, verifiable constraints while keeping prompts expressible in ordinary natural language.

To avoid overlap with LLM training data, each task uses a distinct synthetic compact (\chang{$<10$ columns; 5--365 rows, median 16}) tabular dataset with LLM-generated semantically plausible field names\chang{, following similar strategies in prior visualization benchmarks~\cite{chen2024viseval, wang2024visualization}}. 
We also included a small number of irrelevant noise fields which test whether the model selects the intended fields rather than nearby decoys. 

\parahead{Models}
We ran each task-DSL pair with three models: Gemini 3.1 Pro, GPT-5, and Qwen 3.5 Plus, selected to span different providers and capability tiers (two closed-source, one open-source) and test whether observed patterns are consistent across model families.

\parahead{Test suite}
The final test suite contains  $\textrm{tasks} \times \textrm{DSLs} \times \textrm{models} = 41 \times 10 \times 3 = 1230$ trials, giving the task--DSL--model matrix shown in \figref{fig:teaser}.
We do not view this test suite as a leaderboard for either models or DSLs\chang{, nor do we claim to compare the design superiority of one DSL over another}; instead it is merely an instrument to interrogate the space of possible issues that LLMs can encounter with visualization DSLs. \chang{If LLMs exhibit different error patterns across DSL designs while other factors remain constant, we can reasonably infer that the errors correlate with specific design choices.}
The full test suite (including all Tasks \texttt{T1}--\texttt{T41}) and experiment results are available in our
\href{https://llmgrammar.netlify.app}{\underline{interactive supplement}}.
\chang{They can also be found in a \href{https://github.com/hconhisway/Where-LLM-Fails-Supp}{\underline{github repository}}.}

The evaluation of this test suite has two stages, as in \figref{fig:teaser} center. 
Following prior work~\cite{chen2024viseval, wang2024visualization}, the pipeline checks whether the model output can be parsed as JSON and whether the resulting specification can render an image. 
Second, rendered cases are \chang{evaluated} with an 11-item rubric grouped into four axes: \textit{structure}, \textit{encoding}, \textit{fidelity}, and \textit{style/readability}. \chang{Each rubric item is scored as pass or fail; a case ``passes all rubric items'' only when every item is satisfied.} The rubric is more fine-grained than a single quality score because our goal is to study how JSON-based visualization specifications fail. 
We developed the rubric through pilot results and iterative discussion among the authors. 
Crucially, we treat the scores as diagnostic rather than evaluative: they help locate failures and support cross-case comparison, from which we infer recurring patterns, instead of providing  model or DSL rankings.

Following recent work on LLMs and VLMs as judges~\cite{lee2024prometheus, liu2023g},
we used a vision-language model, \chang{Claude Opus 4.7}, to evaluate the rubric and task requirements for each rendered image. 
A human reviewer, the first author, then manually reviewed and audited the model scores and the generated notes about the failures. 
Reported rubric numbers use the post-audit results\chang{; when a case was reviewed more than once, we use its most recent review}.
We then grouped failures and iteratively identified recurring error patterns across DSLs, chart types, and models. 


\parahead{Limitations}
This experiment has a deliberately narrow scope, and the results should be read with several limitations in mind.
First, our goal is to evaluate factors in DSL design that lead to model usage failures,
rather than evaluate individual model performance.
We also do not cover the full problem of natural-language-to-visualization using LLMs---which has been \chang{studied in depth}~\cite{luo2021natural}---instead focusing solely on the relationship between DSL design and error forms. 
%
Second, the results of our experiment may be affected by prompt engineering---giving the model documentation files, examples, or a BNF grammar description would likely improve some results~\cite{wang2023grammarprompting}.
Adding those conditions would also make the study much more complex, so we use a single, simple prompting style throughout.
\chang{Our zero-shot design is intentional: it measures each model's internalized representation of a DSL---the baseline that documentation, RAG, or grammar-prompting builds upon. Adding documentation would confound DSL design effects with documentation quality, obscuring the signal we aim to study.}
Next, our experiment does not engage with the context or environment in which a chart is created; we assume the errors are at the same level of abstraction as the DSLs, allowing findings to transfer. 
Finally, we did not run repeated trials for each case. The evaluation pipeline requires extensive, laborious human supervision and post-audit review, which would make the additional trials infeasible to review.
\chang{More broadly, prompt style, synthetic data generation, and model selection are all potential confounders. We control for these by holding each factor constant while varying the DSL, but we do not claim to isolate DSL design as the sole cause of failure. The design considerations we propose are best read as empirically grounded directions that await validation in further studies with broader model sets, prompting strategies, and task distributions.}

\section{Results}

\chang{Of the 1230 trials, 7 (all Gemini) failed to return a result due to recurring generation errors, leaving 1223 generated specifications.
Each stage only receives the cases that passed the previous one.
1222 trials (99.3\% of 1230) passed JSON validation, and 964 (78.4\% of 1230) rendered an image; all 123 ChartML outputs were valid JSON, but none rendered.
Only the 964 rendered cases were rubric-reviewed: 416 of them (43.2\% of rendered cases; 33.8\% of all 1230 trials) passed all 11 rubric items (GPT-5: 117/312; Qwen: 127/308; Gemini: 172/344), and the remaining 548 (56.8\% of rendered cases) failed at least one item.}
\chang{A per-DSL and per-model breakdown of these counts is available in the interactive supplement.}
These numbers should be read as a description of the data generated for qualitative analysis and not as a fair comparison among DSLs or models. 
Because some DSLs have broader vocabularies, some tasks intentionally target difficult chart families, and ChartML was included partly to test what happens when an actively maintained schema is still unfamiliar to current models.


After the rubric review, the first author coded the failures and grouped them into themes via affinity diagramming.
These themes capture how prompt requirements, model behavior, and language design together led to incorrect charts, providing initial evidence about where LLM specification-writing becomes brittle.

\begin{figure}[t]
  \centering
  \includegraphics[width=\linewidth]{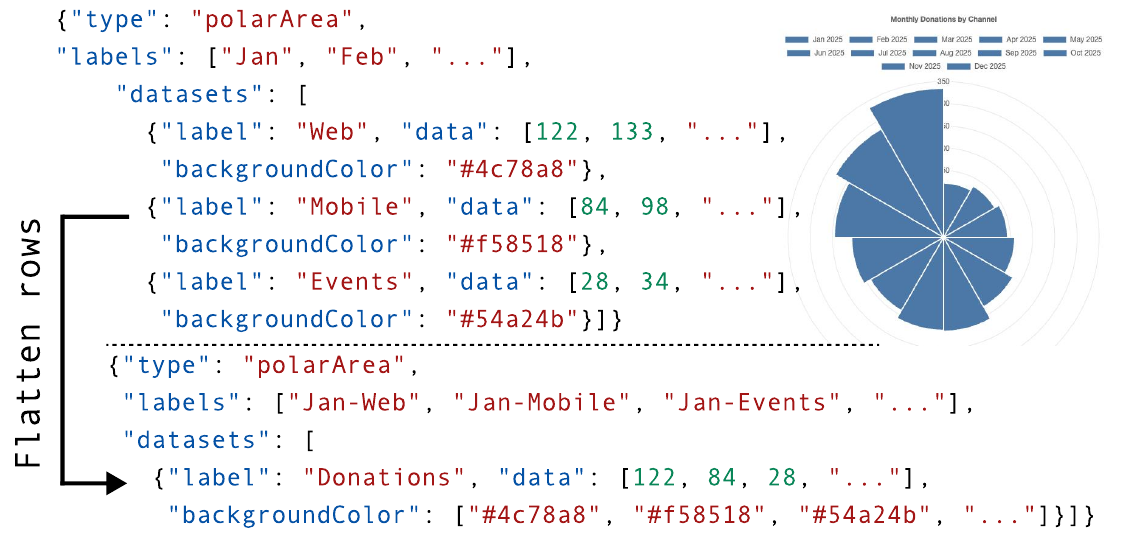}
  \caption{A Chart.js polarArea example where the requested color encoding is entangled with the chart-specific data arrangement, exemplifying the \emph{\chang{Encoding Mismatch}} error. Channel names placed in \texttt{datasets[i].label} do not become wedge colors; the repair flattens month-channel rows and assigns color per wedge.}
  \vspace{-2.5em}
  \label{fig:case1}
\end{figure}

\parahead{\chang{Encoding Mismatch}} 
The first error cluster occurred when models used the right data field in the wrong encoding location.
This happened most often when a DSL did not expose position, value, color, and legend as independent \chang{encoding channels (\ie{} named properties that map a data column to a visual attribute)}, as in Chart.js's \texttt{polarArea}, where labels, values, and colors are distributed across chart-specific arrays.
The output could then look plausible, and even potentially render successfully, while still mis-using a field as a dataset name, node label, color array, or internal trace value rather than as the requested encoding channel (\figref{fig:case1}).

Crucially, the issue is not merely placement of an invalid key. 
Instead, it is that the syntax complects visualization semantics with a chart-specific data shape. 
For example, on Gemini's Chart.js pass of \texttt{T24}: the input table has one row per month-channel pair, with fields for month, donation amount, and channel name. The prompt asks for a rose chart with \texttt{months} around the angle, \texttt{donation\_amount} as the radius, and \texttt{channel\_name} as color and legend. 
Ostensibly, this should be a simple field mapping: month controls position, donation amount controls radius, and channel name controls color. 
However, Chart.js's \texttt{polarArea} does not expose those roles as independent encodings. 
Instead, category labels, legend entries, values, and colors are \emph{inferred} from the chart-specific arrangement of \texttt{data.labels}, \texttt{datasets}, and \texttt{backgroundColor}. 
The generated spec in \figref{fig:case1}
follows a plausible series-based layout by placing months in \texttt{data.labels} and channels in \texttt{datasets[i].label}. For \texttt{polarArea}, that container position has the wrong semantics: \texttt{data.labels} defines the wedge/default-legend categories, while \texttt{datasets[i].label} names a dataset rather than a per-wedge color category. 
The scalar \texttt{backgroundColor} on each dataset likewise colors a whole dataset, not each row. Thus the output is valid JSON, but the encoding is not as requested. 

The same entanglement appears in simpler cases.
In GPT-5's ApexCharts \texttt{T1}, the model stores \texttt{risk\_band} inside each data object and sets \texttt{customLegendItems}---plausible locations, but ApexCharts does not treat them as field-level color encodings. Without \texttt{plotOptions.bar.distributed:true}, bars take the first color rather than varying by category, and the custom labels do not produce a categorical legend. The category is present in the JSON, but not in the DSL location that controls mark color or legend entries.

The observed failures suggest that models are less reliable at translating semantic fields into chart-specific container structures than at filling consistently placed encoding slots, especially when a container position carries different semantics across chart types---echoing McNutt~\cite{mcnutt2022no} this is also likely an issue for humans.


\designcallout{Design Consideration}{
DSLs may be easier for LLMs when common encodings are expressed at the field or channel level. 
Color, legend entries, sort order, and axis titles seem more robust when they can be written as field references, rather than indirectly through the number and names of datasets or series---echoing Cognitive Dimensions of Notation (CDN)~\cite{green1989cognitive},  stronger consistency seems to lead to better performance for humans and models. 
This does not suggest that series-based APIs are unusable, only that  this approach has a greater risk of errors.  
}


\parahead{Layout Arithmetic Failure} Next are instances where the model failed to coordinate spacing, sizing, margins, or panel geometry. 
Besides the data encodings, a range of other features---including multi-panel layout, legends, axis labels, rotated ticks, and titles---all compete for space. 

Plotly's GPT-5 \texttt{T7}  illustrates this well: the title, horizontal legend, and top margin are three separate settings. 
The rendered legend collides with the title because the model puts the legend above the plot but does not allocate enough space for both.

We saw similar layout pressure in larger compositions. 
Vega-Lite facets can overflow the capture frame when the model does not size the cells for a dense grid. 
Vega group layouts can clip panels or shared legends when the model writes per-panel widths without budgeting for outer labels. 
Plotly subplot grids require row spacing, column spacing, trace routing, and panel titles to be coordinated separately. 
These are all slightly different failures, but they share the same basic form: the DSL allows expression of the layout, but it also necessitates several coordinated layout decisions that only become legible after rendering---echoing a similar difficulty in annotations~\cite{rahman2025annogram}. 
\chang{Our results suggest that LLMs may struggle with fine-grained spatial layout decisions, particularly when the correctness of those decisions can only be verified after rendering.} 
This suggests that models are better at declaring compositional intent, such as ``facet by region'' or ``place the legend below,'' than at solving the low-level geometry that some DSLs expose as  parameters.

\designcallout{Design Consideration}{Layout primitives may help if they declaratively allocate space, rather than requiring navigation of multiple implicit constraints. An operator like facet can still expose advanced overrides, but the default path should allocate room for axis labels, titles, panel labels, and shared legends. 
This shifts the model toward declaring structure rather than solving pixel arithmetic---a principle that also supports human usage in GoG-style designs~\cite{satyanarayan2016vega}. }

\parahead{Missed Renderer Defaults} We observed failures related to  omitted settings or missed renderer conventions that fundamentally changed the chart. 
Many errors were not caused by malformed JSON or by choosing the wrong input fields, but instead came from contracts that sit outside the visible part of the generated spec. 
For instance, defaults that engage only when a property is omitted, data shapes that are accepted but interpreted differently from what was intended, strict object forms that silently drop near-misses, and chart-type names whose semantics differ across libraries.

\begin{figure}[t]
  \centering
  \includegraphics[width=\linewidth]{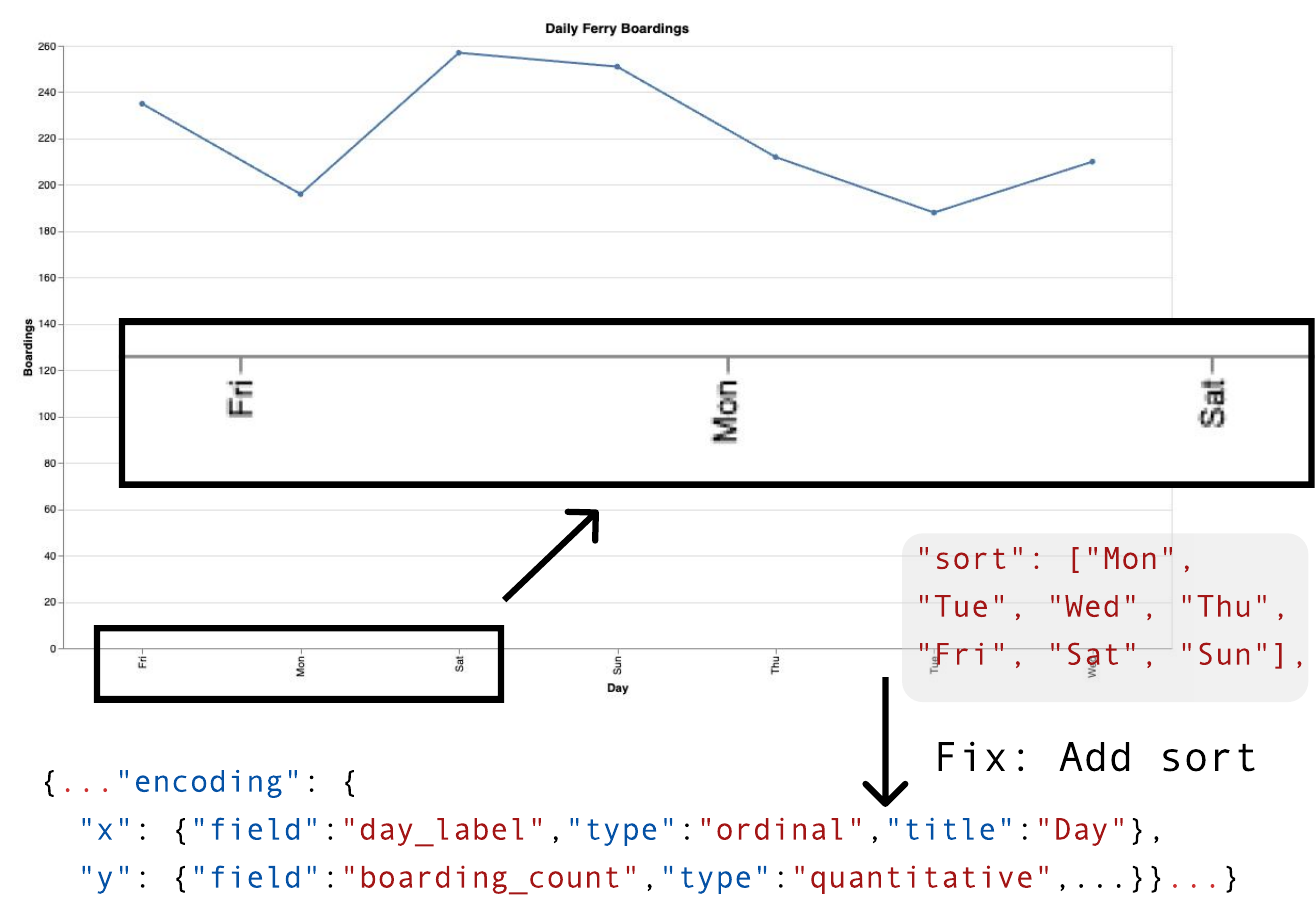}
  \caption{GPT-5's Vega-Lite \texttt{T6} output uses the correct fields for a daily ferry boarding line chart, but omits the weekday order, yielding alphabetized labels.
  }
  \vspace{-2.5em}
  \label{fig:case2}
\end{figure}

A clear default-driven example is shown in \figref{fig:case2}. 
In GPT-5's Vega-Lite \texttt{T6}, the prompt asks for a daily ferry boarding line chart from Monday through Sunday. 
The generated spec chooses the right fields, placing \texttt{day\_label} on the x-axis and \texttt{boarding\_count} on the y-axis, and it uses the requested blue line. 
At the level of field selection, this \emph{looks} correct. The failure comes from the ordering contract attached to the x channel: \texttt{day\_label} is declared as ordinal, but the spec does not provide an explicit \texttt{sort}. 
Vega-Lite then orders the weekday strings alphabetically, yielding days incorrectly sequenced Fri, Mon, Sat, Sun, Thu, Tue, Wed. 

Defaults can add visible content not requested in the task. For instance, GPT-5's ApexCharts \texttt{T12} percentage labels appear inside a donut chart because \texttt{dataLabels.enabled} is set to true. This is not a syntax error, and the chart is recognizable, but the output contains an extra annotation layer relative to the prompt.
Whereas towers of defaults are crucial to the usability of DSLs for humans (\ala{} the design of ggplot2~\cite{wickham2010layered} and Vega-Lite~\cite{satyanarayan2016vega}), this extra implied information can trip up or complicate LLM's DSL use. 

\designcallout{Design Consideration}{Renderers could make more of their hidden behavior visible. A resolved-spec view could show which defaults fired after the first render. Clear parse-time errors could replace silent drops for ignored keys, malformed values, or unsupported data arrangements. Similar chart types could use names and data contracts that are less easy to confuse across libraries. 
These changes would not reduce expressiveness but would make the DSL more legible to models and, as documentation, to humans. }

\parahead{\chang{Failed Workaround}} Lastly were failures where the model used a substitute chart or hand-built workaround that did not preserve the requested semantics.
LLMs tend to return something in response to any request, typically by selecting a nearby primitive or composing lower-level marks.
These attempts are informative because they show which path through the DSL seemed available to the model. For instance, on GPT-5's Vega-Lite \texttt{T19}, the model tries to build a radar chart by layering line and area marks in polar coordinates. The strategy is recognizable, but it does not match Vega-Lite's mark vocabulary: \texttt{theta} and \texttt{radius} are not general polar-position channels for line and area marks. 
The chart therefore renders the title and team legend but almost no radar geometry, collapsing the requested three polygons to an essentially empty plot.
Similarly, on GPT-5's Plotly \texttt{T29}, the model tries to approximate a streamgraph with ordinary stacked area traces using \texttt{stackgroup}.
That recovers the layered area appearance, but fails the defining streamgraph property: the layers are not arranged around a centered baseline.

\begin{wrapfigure}[6]{ri}{0.36\linewidth}
     \vspace{-0.4cm}
       \centering
       \hspace{-8mm}
       \includegraphics[width=1.1\linewidth]{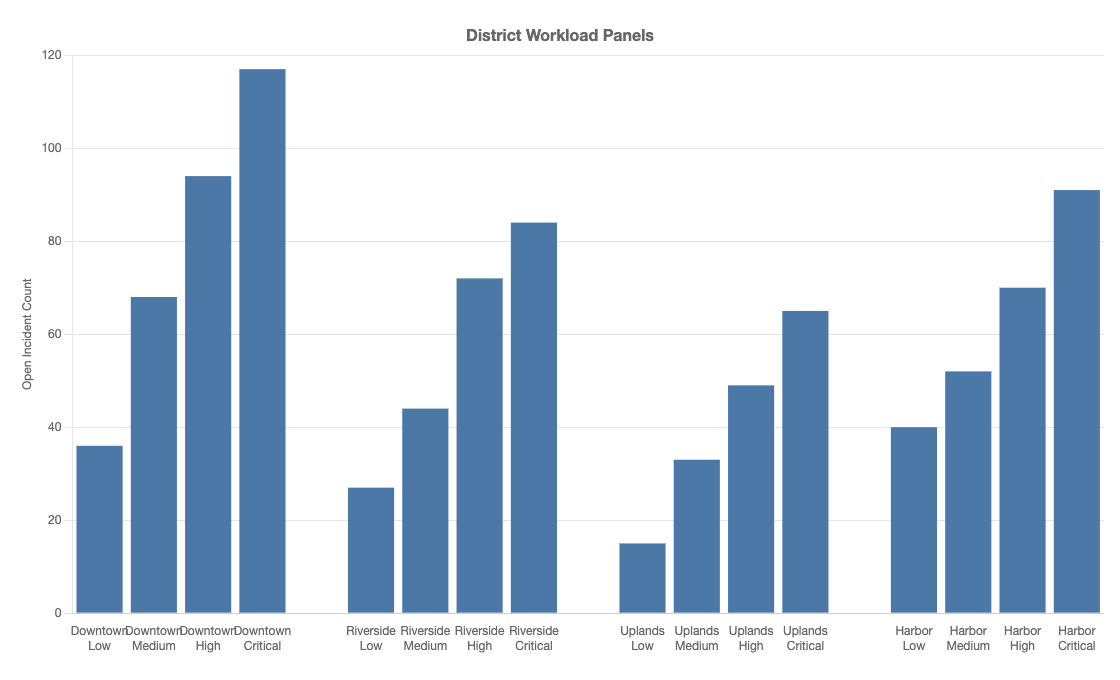}
       \chang{\label{fig:chartjs-t37}}
   \end{wrapfigure}
Not all workarounds fail. For instance, Chart.js has no first-class facet primitive, but on Gemini's Chart.js \texttt{T37}, the model still produces a usable faceted bar chart (\chang{\figref{fig:chartjs-t37}}) by flattening the panels into one bar chart. 
Each x-axis label combines the district and priority, and null data values create visual gaps between district groups, yielding a makeshift small-multiples. 

While the failures above were about the wrong workaround, ChartML's failures are about not knowing the DSL in the first place. 
All 41 cases fail for every model with the same parser error, because the models simply do not know this syntax. 
ChartML is recent and has little public presence, so none of the three models had learned its syntax, which is not supplied in the prompt.
With nothing to go on, each model defaults to the nearest DSL it does know---Vega-Lite-shaped specs from Gemini and Qwen, a Highcharts-style \verb|{chart: {...}}| wrapper from many GPT-5 outputs---none of which match ChartML's actual top-level shape (\texttt{type:~chart}, \texttt{data}, \texttt{visualize}), so the spec is rejected before any design choice in ChartML can be investigated. 
\chang{ChartML controls for unfamiliarity: despite its 100\% parse-failure rate, all four patterns persist in familiar DSLs.}
This, unsurprisingly, confirms that adoption and public examples \chang{are a} prerequisite for LLM authoring without documentation augmentation.

\designcallout{Design Consideration}{Support for less common chart types does not always require a separate named primitive for every chart type. It may be just as important to provide a small set of composition tools that behave predictably. 
Named primitives are useful when the model knows what inputs they expect. But when the DSL  does not include the requested chart directly, the model needs a reliable way to assemble it from simpler parts, which is especially critical for less prominent DSLs.
This is notably different than human behavior, where a human would likely give up or switch tools. 
}

\section*{Acknowledgments}
We used LLM to polish grammar and language.
\vspace{-2mm}

\bibliographystyle{abbrv-doi}

\bibliography{template}

\end{document}